\documentclass[10pt,letterpaper]{article}
\usepackage{opex3}
\usepackage{bm}

\begin{document}

\title{Fourier factorization with complex polarization bases in the plane-wave expansion method applied to two-dimensional photonic crystals$^\star$}

\author{Roman Antos and Martin Veis}

\address{Institute of Physics, Faculty of Mathematics and Physics, Charles University}

\email{antos@karlov.mff.cuni.cz} 



\begin{abstract}
We demonstrate an enhancement of the plane wave expansion method treating two-dimensional photonic crystals by applying Fourier factorization with generally elliptic polarization bases. By studying three examples of periodically arranged cylindrical elements, we compare our approach to the classical Ho method in which the permittivity function is simply expanded without changing coordinates, and to the normal vector method using a normal--tangential polarization transform. The compared calculations clearly show that our approach yields the best convergence properties owing to the complete continuity of our distribution of polarization bases. The presented methodology enables us to study more general systems such as periodic elements with an arbitrary cross-section or devices such as photonic crystal waveguides.\\
$^\star$\textbf{Published in Optics Express 18 [26], 27511--27524 (2010)}
\end{abstract}

\ocis{(050.5298) Photonic crystals; (050.1755) Computational electromag. methods} 



\section{\label{sec:intro}Introduction}

Photonic crystals (PhCs) are modern artificially structured systems of a broad interest from many viewpoints of fundamental research and applications~\cite{Joannopoulos,Yablonovitch_PRL_1987,Ho_PRL_1990,Joannopoulos_Nature_1997}. Studying their electromagnetic properties is significantly important for developing PhC-based promising applications such as purely optical integrated circuits~\cite{Lin_Science_1998,Noda_JLT_2006,Hugonin_OL_2007}, artificial metamaterials with high tunability~\cite{Datta_PRB_1993,Genereux_PRB_2001,Krokhin_PRB_2002,Reyes_PRB_2005}, high-sensitivity PhC biosensors~\cite{Skivesen_OE_2007,Block_IEEESensJ_2008}, or devices based on phenomena not accessible in conventional media~\cite{Kosaka_PRB_1998,Krokhin_PRL_2004,Benisty_JOSAB_2009}. It is also necessary for accounting for structural colors of wings of butterflies or beetles, feathers of birds, or iridescent plants~\cite{Vukusic_Nature_2003,Kinoshita_ChPhCh_2005}.

Owing to the PhC periodicity, the plane-wave expansion (PWE) method is conventional for calculating PhC modes and photonic band structures. Its essence is the Fourier expansion of electromagnetic fields and material parameters, which is not only a counterpart of the PWE method for electronic crystals, but it also advantageously follows the classical coupled-wave theory developed for diffraction gratings during last several decades~\cite{Petit,Neviere,Maystre_PO_1984}. One of the most important discoveries of the grating theory are the Fourier factorization (FF) rules for expanding the ratios of the discontinuous permittivity and the discontinuous electric field~\cite{Li_JOSAA_1996}. The three FF theorems were first formulated for one-dimensional (1D) gratings and then generalized to two-dimensional (2D) systems~\cite{Li_JOSAA_1997}, arbitrary periodic reliefs~\cite{Popov_JOSAA_2000}, anisotropic~\cite{Li_JMO_1998} and slanted~\cite{Chernov_OC_2001} periodic structures, their various combinations~\cite{Watanabe_JOSAA_2002a,Watanabe_JOSAA_2002b,Li_JOA_2003}, and other applications~\cite{Boyer_JOSAA_2004,Bonod_OC_2005a,Bonod_OC_2005b}. As will be demonstrated in this paper, the FF rules establish suitable principles for an accurate solution of Maxwell's equations in the plane-wave basis, which considerably enhances the numerical performance compared to previously used implementations.

In the case of 2D-periodic structures, the FF rules were first applied to ``zigzag'' Fourier expansions~\cite{Li_JOSAA_1997}, which yielded an improvement for rectangular dots or holes, but were not sufficient for the staircase approximation of round elements. After that a coordinate transform was applied to treat individually the normal and tangential components of the electric field on 1D sinusoidal-relief gratings, which enabled the application of the correct rule for each field component~\cite{Popov_JOSAA_2000}. Later \mbox{David et al.}~\cite{David_PRB_2006} utilized the normal--tangential field separation to 2D PhCs composed of circular and elliptical holes. Similarly, Schuster and colleagues~\cite{Schuster_JOSAA_2007} applied this method to 2D gratings, and also suggested more general distributions of polarization bases~\cite{Gotz_OE_2008}. These approaches, always dealing with linear polarizations, enabled a significant improvement of the convergence properties, but ignored the fact that the transformation matrix between the Cartesian and the normal--tangential component bases of polarization became discontinuous at the center and along the boundaries of the periodic cell, which slowed down the resulting convergence. To overcome these discontinuities, a distribution of more complex (i.e., generally elliptic) polarization bases was recently suggested to improve optical simulations of 2D gratings~\cite{Antos_OE_2009}.

Here we are dealing with 2D PhCs made as 2D-periodic elements of the circular cross-section, for which we implement the PWE method by using FF with complex polarization bases. From the mathematical point of view this complexity only requires complex-valued Jones vectors, so that the generalization from the previous method of David is surprisingly simple. Both methods are compared with respect to their convergence properties, together with the classical method of Ho~\cite{Ho_PRL_1990} (where the electric permittivity is expanded without any coordinate transform), to show that the method presented here yields the best performance.

\section{\label{sec:PWE}PWE method for 2D structures}

\begin{figure}
\centerline{\includegraphics[width=11cm]{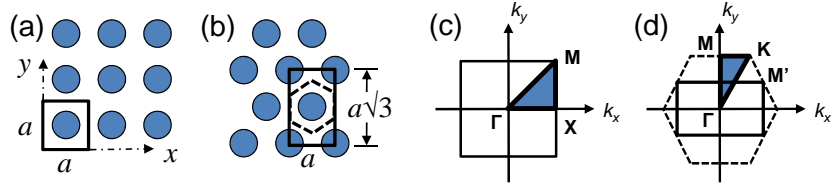}}
\caption{\label{fig:config} Two studied configurations of 2D PhCs, with square (a) and hexagonal (b) symmetry, with denoted square and rectangular unit cells, respectively. The corresponding first Brillouin zones of the reciprocal space are depicted in (c) and (d), respectively, where the emphasized triangles denote irreducible fractions. Note that for calculations we use a rectangular unit supercell (solid rectangle) rather than a hexagonal primitive cell (dashed hexagon) in the case of the hexagonal lattice (b) so that the corresponding first Brillouin zone is the solid rectangle instead of the dashed hexagon in (d), which creates a folded band structure, from where we choose values corresponding to the usual convention.}
\end{figure}

We study a 2D PhC composed of infinite cylinders with a circular cross-section with either square [Fig.~\ref{fig:config}(a)] or hexagonal [Fig.~\ref{fig:config}(b)] periodicity. While the structure is invariant along the $z$-axis, the periodicity is assumed in the $x$- and $y$-axes directions. For the square symmetry the unit cell has the dimensions $a_x=a_y=a$, and for the hexagonal symmetry it can be chosen as an $a$-by-$a\sqrt{3}$ rectangle. The corresponding first Brillouin zones of the reciprocal space are depicted in Figs.~\ref{fig:config}(c) and~\ref{fig:config}(d), together with the symmetry points $\Gamma$, X, M, and $\Gamma$, M, K, respectively. According to this 2D-periodic structure we define and expand the relative permittivity function
\begin{equation}
 \varepsilon(x,y)=\sum_{m,n=-\infty}^{+\infty}\varepsilon_{mn}e^{-i(mG_xx+nG_yy)},
\end{equation}
where $\varepsilon_{mn}$ are its Fourier coefficients and $G_x=2\pi/a_x$ and $G_y=2\pi/a_y$ are the reciprocal lattice vectors.

Maxwell's equations for the $H$-polarization (where the electric field $\mathbf{E}$ is in the $x$--$y$ plane whereas the magnetic field $\mathbf{H}$ is along the $z$-axis) are
\begin{eqnarray}
 &\partial_y H_z = i\omega\varepsilon_0\varepsilon E_x,\\
 &-\partial_x H_z = i\omega\varepsilon_0\varepsilon E_y,\\
 &\partial_x E_y - \partial_y E_x = -i\omega\mu_0 H_z,
\end{eqnarray}
where $\omega$ is the photon frequency, and $\varepsilon_0$ and $\mu_0$ are the permittivity and permeability of vacuum, or
\begin{eqnarray}
 &\left[\begin{array}{lr}-\partial_y,& \partial_x \end{array}\right]  \left[\begin{array}{c}E_x\\ E_y\end{array}\right] = -i\omega\mu_0 H_z, & \label{eq:Maxwell2a}\\
 &\varepsilon\left[\begin{array}{c}E_x\\ E_y\end{array}\right] =
           {\displaystyle \frac{1}{i\omega\varepsilon_0}} \left[\begin{array}{c}\partial_y\\ -\partial_x\end{array}\right]H_z.&\label{eq:Maxwell2b}
\end{eqnarray}
Note that here we do not study the $E$-polarization (for which the electric field $\mathbf{E}$ is along the $z$-axis) because in that case the Ho method satisfies the correct Fourier factorization rules.

Assuming an in-plane anisotropy of the relative permittivity function, defining a scaled electrical displacement $\tilde{\mathbf{D}}$,
\begin{equation}\label{eq:D-field}
 \left[\begin{array}{c}\tilde{D}_x\\ \tilde{D}_y\end{array}\right] = \bm{\varepsilon}\left[\begin{array}{c}E_x\\ E_y\end{array}\right] =
          \left[\begin{array}{ll}\varepsilon_{xx} & \varepsilon_{xy}\\ \varepsilon_{yx} & \varepsilon_{yy} \end{array}\right]
          \left[\begin{array}{c}E_x\\ E_y\end{array}\right],
\end{equation}
defining a 2-by-2 matrix of electrical impermittivity $\bm{\eta}=\bm{\varepsilon}^{-1}$, and substituting Eq.~(\ref{eq:Maxwell2b}) into Eq.~(\ref{eq:Maxwell2a}) yields
\begin{equation}\label{eq:eigen}
 (-\partial_y\eta_{xx}\partial_y + \partial_x\eta_{yx}\partial_y + \partial_y\eta_{xy}\partial_x - \partial_x\eta_{yy}\partial_x)H_z=
       \frac{\omega^2}{c^2}H_z,
\end{equation}
where $\eta_{jk}$ are the components of the electrical impermittivity, $c=(\varepsilon_0\mu_0)^{-1/2}$ is the light speed in vacuum, and the factor $\omega^2/c^2$ is calculated as the eigenvalues of the operator on the left-hand side of Eq.~(\ref{eq:eigen}).

According to the Floquet--Bloch theorem, in the reciprocal space the magnetic field undergoes the Fourier expansion
\begin{equation}
 H_z(x,y)=\sum_{m,n=-\infty}^{+\infty}h_{z,mn}e^{-i(p_mx+q_ny)},
\end{equation}
where $p_m=k_x+mG_x$, $q_n=k_y+nG_y$. Analogously, the Fourier components of $E_x$, $E_y$, $\tilde{D}_x$, $\tilde{D}_y$ are denoted $e_{x,mn}$, $e_{y,mn}$, $\tilde{d}_{x,mn}$, $\tilde{d}_{y,mn}$. Then, as demonstrated in Appendix, the operators of the partial derivatives $\partial_x$, $\partial_y$ become the diagonal matrices $-i\mathbf{p}$, $-i\mathbf{q}$, the operators of multiplication by functions such as $\varepsilon_{xx}$ or $\eta_{xx}$ become matrices denoted $[\![\varepsilon_{xx}]\!]$ or $[\![\eta_{xx}]\!]$, and the field components $E_x$, $E_y$, $H_z$ become column vectors denoted $[E_x]$, $[E_y]$, $[H_z]$. Using these definitions, the multiplications in Eq.~(\ref{eq:D-field}) can be treated either by the Laurent FF rule, $[\varepsilon_{xx}E_{x}]=[\![\varepsilon_{xx}]\!][E_x]$ etc., or by the inverse FF rule, $[\varepsilon_{xx}E_{x}]=[\![1/\varepsilon_{xx}]\!]^{-1}[E_x]$.

\section{\label{sec:FF}Methods of Fourier factorization}

\subsection{Elementary (Cartesian) method (Model~A)}

\begin{figure}
\centerline{\includegraphics[width=13cm]{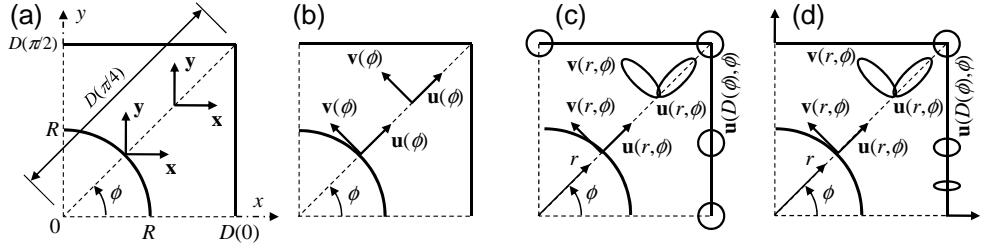}}
\caption{\label{fig:polarization} Schematic description of the polarization distributions in Models A~(a), B~(b), C~(c), and C'~(d) within the first quadrant of the periodic cell, as functions of the polar coordinates $re^{i\phi}=x+iy$. In (a) the $\mathbf{x}$--$\mathbf{y}$ Cartesian basis is uniform; in (b) the polarization vectors are normal ($\mathbf{u}$) and tangential ($\mathbf{v}$) to the cylindrical element and constant along lines coming from the center; in (c) and (d) the $\mathbf{u}$, $\mathbf{v}$ polarizations are in general elliptic which enables their continuity.}
\end{figure}

In this article we compare several models corresponding to different FF approaches. First, Model~A assumes the solution in the basis of the $\mathbf{x}$ and $\mathbf{y}$ polarizations uniform within the periodic cell [Fig.~\ref{fig:polarization}(a)], where in accordance with the Ho method~\cite{Ho_PRL_1990} we choose the Laurent rules
\begin{eqnarray}
 &{[}\tilde{D}_x] = [\varepsilon E_x] = [\![ \varepsilon ]\!][E_x],\label{eq:Laurent1}&\\
 &{[}\tilde{D}_y] = [\varepsilon E_y] = [\![ \varepsilon ]\!][E_y].\label{eq:Laurent2}&
\end{eqnarray}
The components of the electric impermittivity in Eq.~(\ref{eq:eigen}) then becomes $[\![\varepsilon]\!]^{-1}$ for the cases of $\eta_{xx}$, $\eta_{yy}$, and zero for the cases of $\eta_{xy}$, $\eta_{yx}$, or
\begin{equation}
 [\![\bm{\eta}]\!]_{\rm A} = \left[ \begin{array}{ll}[\![ \varepsilon ]\!]^{-1} & {[\![0]\!]} \\ {[\![0]\!]} & [\![ \varepsilon ]\!]^{-1} \end{array}\right].
\end{equation}
In Fig.~\ref{fig:polarization}(a) we depict the uniform polarization basis within the first quadrant of the square periodic cell, $0<x<D(0)$, $0<y<D(\pi/2)$, where $D(\phi)$ is the distance between the cell's center and its boundary (a function of the polar coordinate~$\phi$). We also denote $R$ the radius of the cylindrical element.

\subsection{Normal vector method (Model~B)}

According to Li~\cite{Li_JOSAA_1996} the Laurent rule is valid (for the reason of uniform convergence) for multiplying functions that possess no concurrent discontinuities, whereas the inverse rule is used for functions whose product is continuous. Obviously, neither the Laurent rule nor the inverse rule is correct for both products in Eqs.~(\ref{eq:Laurent1}) and~(\ref{eq:Laurent2}), because both pairs of functions have concurrent discontinuities and both products $\tilde{D}_x$ and $\tilde{D}_y$ are discontinuous as well. On the other hand, by an appropriate change of the polarization bases at all points (using a space-dependent Jones matrix transform $\mathbf{F}$),
\begin{equation}
 \left[\begin{array}{l}E_x\\E_y\end{array}\right] = \mathbf{F} \left[\begin{array}{l}E_u\\E_v\end{array}\right],
\end{equation}
we can treat independently the normal ($u$) and tangential ($v$) components of the fields by the correct rules,
\begin{eqnarray}
 {[}\tilde{D}_u]&=&[\![1/\varepsilon]\!]^{-1}[E_u],\label{eq:Du1}\\
 {[}\tilde{D}_v]&=&[\![\varepsilon]\!][E_v].\label{eq:Dv1}
\end{eqnarray}
The field components $E_u$, $\tilde{D}_u$ are normal to the discontinuities of the relative permittivity function, while $E_v$, $\tilde{D}_v$ are tangential. The FF rules used in Eqs.~(\ref{eq:Du1}) and~(\ref{eq:Dv1}) are justified simply because $E_v$ and $\tilde{D}_u$ are continuous.

A suitable distribution of the matrix $\mathbf{F}$ within the periodic cell can obviously be the rotation~\cite{David_PRB_2006}
\begin{equation}
 \mathbf{F}=\left[\begin{array}{cc}\cos\phi & -\sin\phi\\ \sin\phi & \cos\phi\end{array}\right],
\end{equation}
where the polar angle $\phi(x,y)$ is in the first cell distributed according to the polar coordinates $re^{i\phi}=x+iy$, and then periodically repeated over the entire 2D space. This enables defining the matrices $[\![c]\!]$, $[\![s]\!]$ from the corresponding 2D-periodic functions $c=\cos\phi$, $s=\sin\phi$.

\begin{figure}
\centerline{\includegraphics[width=8cm]{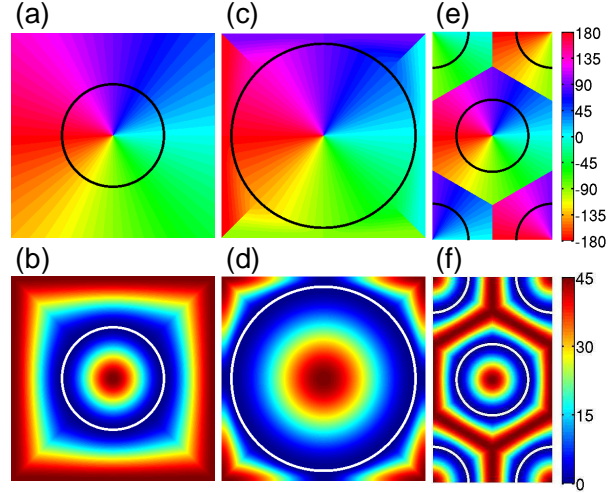}}
\caption{\label{fig:cells} Distribution of the rotation and ellipticity of the basis polarization vector $\mathbf{u}$ for the presented models. The rotation for Models~B and~C is in (a) [(e) for the hexagonal periodicity]; the ellipticity for Model~C is in~(b) [(f) for the hexagonal periodicity]; and the rotation and ellipticity for Model~C' are in (c) and~(d), respectively. The color scales for both the rotation and the ellipticity are on the right (in degrees). The structure discontinuities (the circular boundaries of the periodic elements) are plotted as black or white circles. Notice that $\mathbf{u}$ is always linear along and normal to the circle.}
\end{figure}

Let $\mathbf{u}$ and $\mathbf{v}$ be the two columns of the matrix $\mathbf{F}$, both being mutually orthogonal basis vectors of linear polarization~\cite{Azzam}. From the above definitions we see that $\mathbf{u}$ is a polarization vector normal to the structure discontinuities, whereas $\mathbf{v}$ is tangential. In Fig.~\ref{fig:polarization}(b) we depict the distribution of the $\mathbf{u}$--$\mathbf{v}$ basis within the first quadrant of the periodic cell; the polarization vectors are constant along the lines of the constant azimuth ($\phi={\rm const}$) and rotate as $\phi$ increases. The distribution of the rotation of $\mathbf{u}$ within the square periodic cell is displayed in Fig.~\ref{fig:cells}(a), from where it is obvious that the matrix function $\mathbf{F}(x,y)$ has no discontinuities concurrent with the electric field, so that we can use both Laurent and inverse rules for the transformation of polarization, e.g.,
\begin{eqnarray}\label{eq:F_transform}
&\left[\begin{array}{c}{[E_x]}\\{[E_y]}\end{array}\right] = [\![\mathbf{F}]\!]\left[\begin{array}{c}{[E_u]}\\ {[E_v]}\end{array}\right],&\\
&{[\![}\mathbf{F}]\!] = \left[\begin{array}{cc}{[\![c]\!]} & {[\![-s]\!]}\\ {[\![s]\!]} & {[\![c]\!]}\end{array}\right].&
\end{eqnarray}
Combining Eqs.~(\ref{eq:Du1}), (\ref{eq:Dv1}), and (\ref{eq:F_transform}) yields
\begin{equation}\label{eq:impermittivity}
\left[\begin{array}{c}{[E_x]}\\{[E_y]}\end{array}\right] = [\![\mathbf{F}]\!]
\left[\begin{array}{ll} {[\![\frac{1}{\varepsilon}]\!]} & {[\![0]\!]}\\ {[\![0]\!]} & {[\![\varepsilon]\!]^{-1}}\end{array}\right]
{[\![\mathbf{F}^{-1}]\!]} \left[\begin{array}{c}{[\tilde{D}_x]}\\{[\tilde{D}_y]}\end{array}\right],
\end{equation}
from where we derive the electric impermittivity in the reciprocal space (corresponding to Model~B)
\begin{eqnarray}
[\![\bm{\eta}]\!]_{\rm B} &=& [\![\mathbf{F}]\!]
\left[\begin{array}{ll} {[\![\frac{1}{\varepsilon}]\!]} & {[\![0]\!]}\\ {[\![0]\!]} & {[\![\varepsilon]\!]^{-1}}\end{array}\right]
{[\![\mathbf{F}^{-1}]\!]}  \nonumber \\
&=& \left[ \begin{array}{lr}
{}[\![\frac{1}{\varepsilon}]\!][\![c^2]\!]+[\![\varepsilon]\!]^{-1}[\![s^2]\!], & [\![\frac{1}{\varepsilon}]\!][\![cs]\!]-[\![\varepsilon]\!]^{-1}[\![cs]\!]\\
{}[\![\frac{1}{\varepsilon}]\!][\![cs]\!]-[\![\varepsilon]\!]^{-1}[\![cs]\!], & [\![\frac{1}{\varepsilon}]\!][\![s^2]\!]+[\![\varepsilon]\!]^{-1}[\![c^2]\!]
\end{array} \right],
\label{eq:etaB_detail}
\end{eqnarray}
whose components are immediately applicable to Eq.~(\ref{eq:eigen}).

\subsection{Method with elliptical polarization bases (Model~C)}

The above approach (Model~B) only deals with linear polarizations and thus suffers from the fact that the matrix function $\mathbf{F}(x,y)$ has a singularity at the center of the periodic cell and other discontinuities along the cell boundaries. This slows down the convergence of the numerical implementation, as will be evidenced below.

As suggested in the previous work for the case of diffraction by 2D gratings~\cite{Antos_OE_2009}, we can make $\mathbf{F}$ continuous by using complex functions $\xi$ and $\zeta$ or, in other words, by defining $\mathbf{u}$,~$\mathbf{v}$ as complex vectors corresponding to generally elliptic polarizations,
\begin{equation}\label{eq:uv}
 \mathbf{u}=\left[\begin{array}{c}\xi\\ \zeta \end{array}\right],\quad
 \mathbf{v}=\left[\begin{array}{c}-\zeta^\ast\\ \xi^\ast \end{array}\right]
\end{equation}
(which are still orthogonal). By means of rotation~$\theta$ and ellipticity~$\mathcal{E}$ we define the first basis vector
\begin{equation}\label{eq:CFF:u1}
 \mathbf{u}=e^{i\theta}\left[\begin{array}{cc}\cos\theta & -\sin\theta\\ \sin\theta & \cos\theta \end{array}\right]
                       \left[\begin{array}{c}\cos\mathcal{E} \\ i\sin\mathcal{E} \end{array}\right],
\end{equation}
where
\begin{eqnarray}
\theta(r,\phi) &=& \phi,\label{eq:CFF:azimuth}\\
\mathcal{E}(r,\phi) &=& \left\{
\begin{array}{ll}
\frac{\pi}{8}\left(1+\cos\frac{\pi r}{R}\right) & (r\le R)\\
\frac{\pi}{8}\left\{1+\cos\frac{\pi[r+D(\phi)-2R]}{D(\phi)-R}\right\} & (r>R).
\end{array}\right.\label{eq:CFF:ellipticity}
\end{eqnarray}
Here $R$ denotes the radius of the circular element and
\begin{equation}\label{eq:D1}
 D(\phi)=\frac{a/2}{\max(|\cos\phi|,|\sin\phi|)}
\end{equation}
is the distance from the cell's center to its edge. In Eq.~(\ref{eq:CFF:u1}) the Jones vector on the right represents a polarization ellipse (with ellipticity~$\mathcal{E}$) oriented along the $x$ coordinate, the matrix in the middle rotates this polarization by the azimuth~$\theta$, and the factor~$e^{i\theta}$ preserves the continuity of the phase at the center and along the boundaries of the cell. This continuity can be easily checked by evaluating the limits
\begin{equation}
\lim_{r\to 0} \mathbf{u} = \lim_{r\to D(\phi)} \mathbf{u} = \frac{1}{\sqrt{2}}\left[\begin{array}{c} 1\\i \end{array}\right],
\end{equation}
which is the vector of left circular polarization (independent of~$\phi$).

The distribution of the {$\mathbf{u}$--$\mathbf{v}$} polarization basis within the first quadrant of the periodic cell is schematically depicted in Fig.~\ref{fig:polarization}(c), where the azimuth of the polarization ellipse is constant along the lines coming from the cell's center, which is similar to Model~B; hence, Fig.~\ref{fig:cells}(a) serves for both Model~B and Model~C. However, the ellipticity is now zero (corresponding to linear polarization) only on the boundaries of the circular element, has the maximum value ($\pi/4$ for circular polarization) at the cell's center and along its boundaries, and continuously varies (with a smooth sine dependence) in the intermediate ranges [Fig.~\ref{fig:cells}(b)]. Finally we obtain a smooth and completely continuous matrix function $\mathbf{F}(x,y)$, which is analogously used to calculate the impermittivity in the reciprocal space
\begin{equation}\label{eq:etaC}
[\![\bm{\eta}]\!]_{\rm C} = \left[ \begin{array}{lr}
{[\![\frac{1}{\varepsilon}]\!]}[\![\xi\xi^\ast]\!]+{[\![\varepsilon]\!]}^{-1}[\![\zeta\zeta^\ast]\!], & {[\![\frac{1}{\varepsilon}]\!]}[\![\xi\zeta^\ast]\!]-{[\![\varepsilon]\!]}^{-1}[\![\xi\zeta^\ast]\!]\\
{[\![\frac{1}{\varepsilon}]\!]}[\![\xi^\ast\zeta]\!]-{[\![\varepsilon]\!]}^{-1}[\![\xi^\ast\zeta]\!], & {[\![\frac{1}{\varepsilon}]\!]}[\![\zeta\zeta^\ast]\!]+{[\![\varepsilon]\!]}^{-1}[\![\xi\xi^\ast]\!]
\end{array} \right].
\end{equation}

In the case of the hexagonal periodicity we define $\mathbf{u}$ and the other periodic quantities inside one hexagon (half the area of the rectangular unit cell) where we can use formally the same equations as above, except for
\begin{equation}\label{eq:D2}
 D(\phi)=\frac{a/2}{{\displaystyle\max_{n=0,...,5}}\left[\cos\left(\phi-\frac{n\pi}{3}\right)\right]},
\end{equation}
which is now the distance from the hexagon's center to its edge. Here $a$ is the hexagon's shortest width (equal to the width $a_x$ of the rectangular cell). After periodic repeating over the entire 2D space, the azimuth and ellipticity distributions of the basis vector $\mathbf{u}$ become those displayed in Figs.~\ref{fig:cells}(e) and~\ref{fig:cells}(f), respectively. Note that although the color distribution in Fig.~\ref{fig:cells}(e) is discontinuous along the hexagon's boundaries, the vector $\mathbf{u}$ is actually continuous, because the color difference is only due to the $180^\circ$ change. Moreover, the vector $\mathbf{u}$ in the limit $r\rightarrow D(\phi)$ corresponds to left circular polarization, so that its azimuth becomes irrelevant.

\subsection{Modified method for densely arranged elements (Model~C')}
To analyze a more complicated situation, we consider a PhC with square periodicity where circular elements are densely arranged near each other, i.e., where the radius $R$ is almost the half width $a/2$ of the periodic cell. Then the convergence properties of $\mathbf{F}$ becomes worse, which affects all the derived quantities. For this reason we again redefine the polarization distribution, as suggested in Fig.~\ref{fig:polarization}(d). For the modified Model~C' we define $\mathbf{u}$ to be still same inside the circle ($r<R$), but different outside. Assuming the rotation and ellipticity along the boundary of the square cell
\begin{eqnarray}
 &&\theta_{\rm b}(\phi)=\theta\left(D(\phi),\phi\right) = {\textstyle\frac{\pi}{2}\mathbf{round}\left(\phi/\frac{\pi}{2} \right)},\\
 &&\mathcal{E}_{\rm b}(\phi)=\mathcal{E}\left(D(\phi),\phi\right) = {\textstyle\frac{\pi}{8}(1 - \cos4\phi)}
\end{eqnarray}
(where ``$\mathbf{round}$'' denotes rounding towards the nearest integer), we define the rotation and ellipticity outside the circle ($r>R$) as
\begin{eqnarray}
 \theta(r,\phi)&=&{\textstyle\frac{1}{2}} \left\{ \theta_{\rm b}(\phi)+\phi + [\theta_{\rm b}(\phi)-\phi]\cos {\textstyle\frac{\pi[r+D(\phi)-2R]} {D(\phi)-R}} \right\},\\
 \mathcal{E}(r,\phi)&=& {\textstyle\frac{\mathcal{E}_{\rm b}(\phi)}{2}} \left\{ 1+\cos{\textstyle\frac{\pi[r+D(\phi)-2R]}{D(\phi)-R}} \right\}.
\end{eqnarray}
Assuming otherwise the same Eqs.~(\ref{eq:impermittivity}), (\ref{eq:uv}), (\ref{eq:CFF:u1}), and (\ref{eq:D1}), we obtain for $[\![\bm{\eta}]\!]_{\rm C'}$ formally the same matrix as in Eq.~(\ref{eq:etaC}), except that the functions $\xi$ and~$\zeta$ are now derived from different azimuth and ellipticity distributions of~$\mathbf{u}$ [displayed in Figs.~\ref{fig:cells}(c) and~\ref{fig:cells}(d), respectively]. Note that $\mathbf{u}$ is again continuous along the cell's boundaries; to evaluate its precise limits [when $x\rightarrow\pm D(0)$, $y={\rm const}$ or $y\rightarrow\pm D(\phi/2)$, $x={\rm const}$] is now more complicated.

\begin{figure}
\centerline{\includegraphics[width=13.3cm]{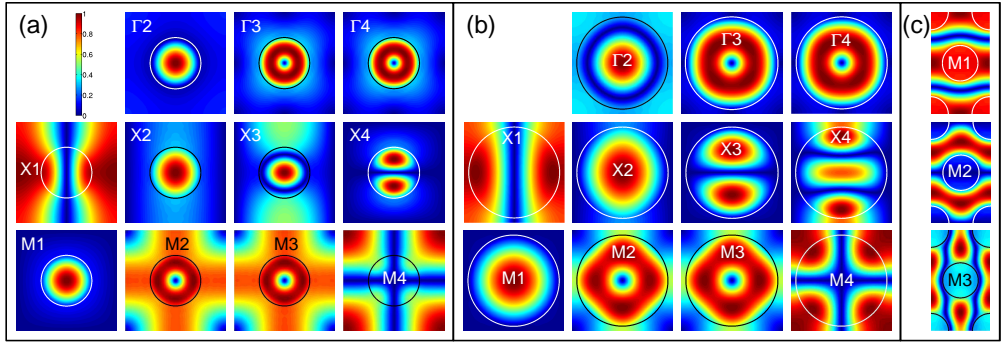}}
\caption{\label{fig:amplitudes} Amplitude distribution of the scaled magnetic field $|H_z(x,y)|$ within one cell of the real space for Samples S1 (a), S2 (b) and~H~(c). The color scale, same for each sample, is in the top left corner of (a). The subfigures represent eigenmodes distinguished by the symmetry point letters ($\Gamma$,~X,~M) and band numbers (1--4). The structure discontinuities (boundaries of the rods or holes) are plotted as white or black circles.}
\end{figure}

\section{Numerical examples}

\subsection{Description of simulations}

We examine the numerical performances of all the presented models on three samples of 2D PhCs, for which we calculate the eigenfrequencies $\omega_\kappa$ (where the band number $\kappa=1$ stands for the lowest eigenfrequency, $\kappa=2$ for the second lowest, etc.) and the corresponding eigenvectors $[H_z]_\kappa$ of Eq.~(\ref{eq:eigen}). All convergences will be presented according to the maximum Fourier harmonics retained inside the periodic medium ($M$, $N$ defined in Appendix), which will be kept same for the $x$ and $y$ directions ($M=N$).

First, Sample~S1 is a square array of cylindrical rods of the circular cross-section with the diameter $2R=500$~nm, square period $a=1000$~nm, relative permittivity of the rods $\varepsilon_1=9$, and relative permittivity of the surrounding medium corresponding to vacuum ($\varepsilon_2=1$). For clarity we display the first several eigenmodes at the points of symmetry $\Gamma$, X, M in Fig.~\ref{fig:amplitudes}(a). The color scale of the modes corresponds to the scaled amplitude distribution of the magnetic field $|H_z(x,y)|$ within one cell of the real space (calculated by Model~C with $M=N=12$). Each eigenmode is denoted by the symmetry point letter and the band number (e.g., X2 denotes the second band at the point~X). By comparing the calculated eigenfrequencies and the amplitude distributions we find that the pairs of eigenmodes $\Gamma3$--$\Gamma4$ and M2--M3 constitute frequency doublets, i.e., the frequencies for one pair are very close to each other (differences are due to numerical errors). Note that although the amplitudes $|H_z(x,y)|$ of the two eigenmodes within a doublet seem identical [e.g., the doublet $\Gamma3$--$\Gamma4$ in Fig.~\ref{fig:amplitudes}(a)], the two actual eigenmodes, which are complex-valued distributions $H_z(x,y)$, are indeed different, being two linearly independent eigenvectors of a degenerate eigenvalue.

Similarly, for Sample~S2 we assume exactly the same parameters except the diameter of the rods, now being $2R=900$~nm. This corresponds to densely arranged elements (the distance between two adjacent rods is only 100~nm). The amplitude distributions for the same bands at the same symmetry points are displayed in Fig.~\ref{fig:amplitudes}(b) (calculated by Model~C' with $M=N=12$), where we again observe the frequency doublets in the same bands.

Finally, for Sample~H we consider a hexagonal array of cylindrical holes of the circular cross-section with the diameter $2R=600$~nm, hexagonal periodicity $a=1000$~nm (corresponding to the rectangular cell of the dimensions $a_x=1~\mu{\rm m}$, $a_y=\sqrt{3}~\mu{\rm m}$), relative permittivity of the holes corresponding to vacuum ($\varepsilon_1=1$), and relative permittivity of the substrate medium (surrounding holes) $\varepsilon_2=12$. Since the lowest (nonzero) eigenfrequencies at the $\Gamma$ and K points are quite high, we confine ourselves only to the first three eigenmodes at the point~M, whose amplitude distributions (within the rectangular cell) are displayed in Fig.~\ref{fig:amplitudes}(c).

\begin{figure}
\centerline{\includegraphics[width=13cm]{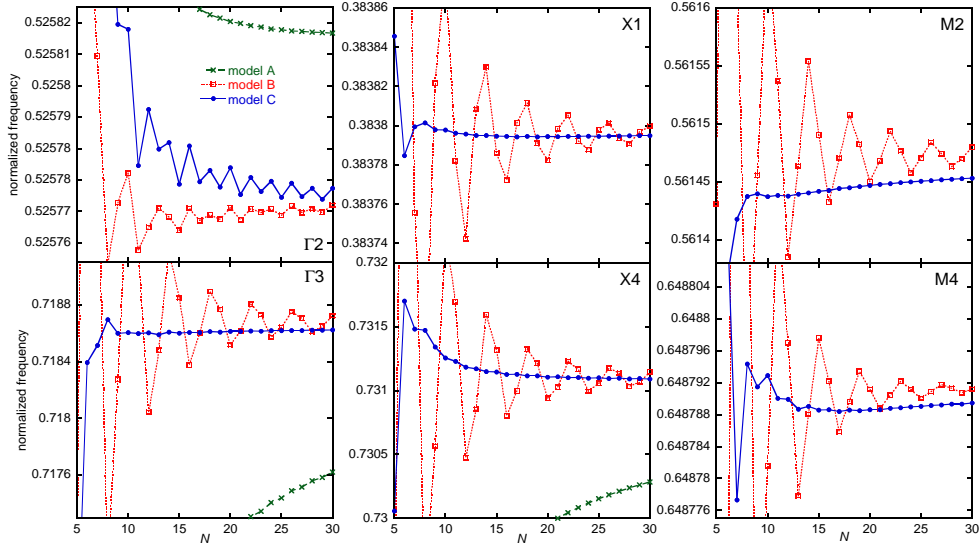}}
\caption{\label{fig:conv_square} Convergence properties of normalized eigenfrequencies ($\omega a/2\pi c$) calculated for selected bands of Sample~S1. The modes $\Gamma2$, $\Gamma3$ are on the left part of the figure; X1, X4 in the middle; and M2, M4 on the right. The Models A, B, and~C are compared to each other, plotted as crosses, squares, and circles, respectively.}
\end{figure}

\subsection{Results and discussion}

We carefully compared the numerical effectivity of the presented models for all the defined samples, and below we present the convergence results for selected bands.

For Sample~S1 the difference among the performances of Models A, B, and~C was found as follows: The values of Model~B always converge much faster than those of Model~A, which is a result expected from previous reports~\cite{David_PRB_2006}. The values of Model~C converge significantly faster than those of Model~B for the modes $\Gamma3$, $\Gamma4$, X1, X4, and M2--M4. This can be explained by the discontinuities of the matrix function $\mathbf{F}(x,y)$ along the boundaries and at the center of the periodic cell, which are responsible for the Gibbs phenomenon appearing in the partial Fourier sums in Eq.~(\ref{eq:etaB_detail}). On the other hand, both Models B and~C exhibit similar convergence properties for the modes $\Gamma2$, X2, X3, and~M1, where the amplitude distribution has almost circular symmetry and is nearly zero near the cell's boundary [Fig.~\ref{fig:amplitudes}(a)], so that the discontinuities do not manifest themselves.

A few examples of compared convergences are displayed in Fig.~\ref{fig:conv_square}, where a considerable improvement owing to Model~C is visible everywhere except for the $\Gamma2$ mode. Nevertheless, according to the scale of the frequency axis at the mode $\Gamma2$ no improvement by Model~C is necessary since Model~B already yields the 5-digits precision. For the modes $\Gamma3$, X1, and X4 the precision is improved by at least one order.

\begin{figure}
\centerline{\includegraphics[width=13cm]{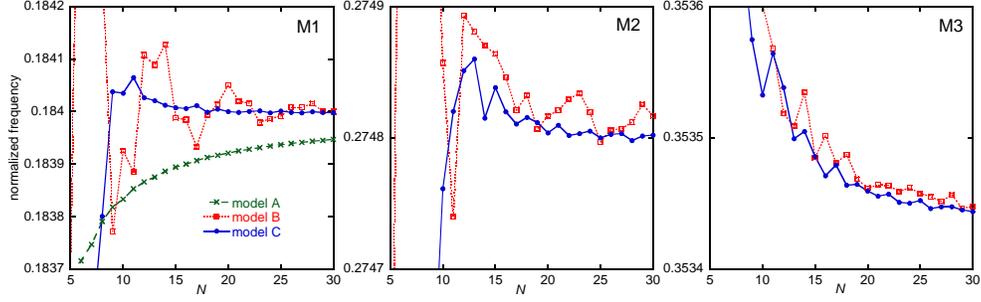}}
\caption{\label{fig:conv_hex} Convergence properties of normalized eigenfrequencies ($\omega a/2\pi c$) calculated for the modes M1 (left), M2 (middle), and M3 (right) of Sample~H. The Models A, B, and~C are plotted as crosses, squares, and circles, respectively.}
\end{figure}

For Sample~H the convergence of Model~A is also much slower than those of Models B and~C. In Fig.~\ref{fig:conv_hex} the values of Model~A are even beyond the displayed range of the graphs for the modes M2 and~M3. Model~C yields the best results for the modes M1 and~M2. On the other hand, Models B and~C converge similarly for the mode~M3; obviously a higher range of~$N$ would be required to compare the convergences more adequately.

All the eigenmodes $\Gamma2$--$\Gamma4$, X1--X4, and M1--M4 of Sample~S2 were carefully analyzed with respect to the convergence properties of Models A, B, C, and C'. Here the values of Model~C', as expected, converge fastest for all the modes except M1, whereas Model~A is again the worst. The eigenmode of the mode M1 is again circularly symmetric and has negligible values near the cell's boundaries [Fig.~\ref{fig:amplitudes}(b)], so that Model~B is not affected by the discontinuities of the polarization transform and yields precise values. On the other hand, Model~C is found inappropriate for this sample (yielding even worse convergence than Model~B) because the rapid variation of the ellipticity of $\mathbf{u}$ near the cell boundaries between two adjacent elements (which are very close to each other) requires more Fourier components than the weak discontinuity of the linear polarization $\mathbf{u}$ in Model~B. This problem is removed in Model~C', as obvious from Figs.~\ref{fig:polarization}(d) and~\ref{fig:cells}(d).

\begin{figure}
\centerline{\includegraphics[width=13cm]{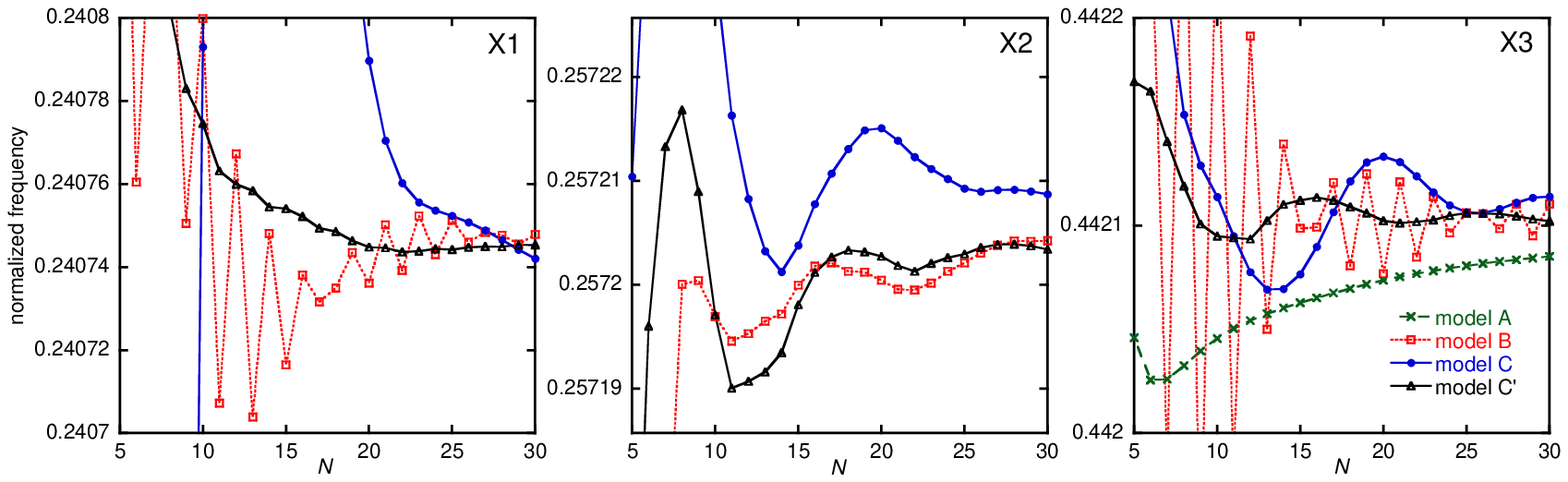}}
\caption{\label{fig:conv_dense} Convergence properties of normalized eigenfrequencies ($\omega a/2\pi c$) calculated for the modes X1 (left), X2 (middle), and X3 (right) of Sample~S2. The Models A, B, C, and~C' are plotted as crosses, squares, circles, and triangles, respectively.}
\end{figure}

Three examples of convergences (for the modes X1--X3) are shown in Fig.~\ref{fig:conv_dense}, which clearly justifies the suitability of Model~C'. Notice again that the values of Model~A are in two cases (modes X1 and X2) beyond the displayed range of the graphs (with error higher by about two orders). In the case of the mode~X3 the values of Model~A do not differ so much, which can be explained by nearly zero values of the eigenmode~X3 at all points of PhC discontinuities, which is obviously not the case of the modes X1 and X2 [Fig.~\ref{fig:amplitudes}(b)].

\section{Conclusions}
We have derived three models with respect to different methods of applying the FF rules and compared their numerical performances. Model~A (equal to the Ho method) exhibits the worst convergence properties. Model~B gives the best performance in a few cases where the eigenmode is circularly symmetric and has a negligible amplitude near the cell's boundaries (then the problems coupled with polarization discontinuities vanish in zero fields). Model~C was found to converge significantly faster than Model~B for PhCs without dense arrangement of elements. On the other hand, for dense PhCs, where Model~C was found inappropriate, the best results were yielded by Model~C', which combines the best properties of Models B and~C. We can conclude that the method of FF with complex polarization bases, represented here by Models C and~C', produces an enhancement of the PWE method, provided that we choose an appropriate distribution of the polarization basis according to the sample under study.

The method can be straightforwardly generalized to PhCs composed of elements of an arbitrary cross-section. It is possible by replacing the constant value of the element's radius~$R$ with a function~$R(\phi)$, and by a further generalization of the azimuth $\theta(r,\phi)$ and ellipticity $\mathcal{E}(r,\phi)$ of the vector~$\mathbf{u}$ to make it again normal to and linear at the element boundaries (and continuous in the entire 2D space). Moreover, the method can also be used to PhC-based devices such as PhC waveguides by applying the demonstrated methodology to the device supercell (similarly as we were here dealing with the rectangular cell of the hexagonal PhC). It is particularly advantageous for devices where high accuracy is required, e.g., for analyzing defect modes near photonic band edges~\cite{Magmoodian_OE_2009,Magmoodian_PRA_2009,Dossou_PRA_2009}, and for large devices for which the available computer memory enables calculations with only a few Fourier harmonics, e.g., PhC-based fibers with large cladding.

\section*{Acknowledgments}
This work is part of the research plan MSM 0021620834 financed by the Ministry of Education of the Czech Republic and was partially supported by a Marie Curie International Reintegration Grant (no.~224944) within the 7th European Community Framework Programme and by the Grant Agency of the Czech Republic (no.~202/09/P355 and P204/10/P346). One of the authors (R.A.) thanks Koki Watanabe and Mathias Vanwolleghem for fruitful discussions.

\appendix
\section*{Appendix: Matrices for 2D structures}
\setcounter{section}{1}

Here we follow the matrix formation procedures and notation as explained by Visnovsky and Yasumoto~\cite{Visnovsky_CJP_2001}. Let $f(x,y)$ be one component of the pseudo-periodic electric field inside the medium described by the 2D-periodic function $\varepsilon(x,y)$. Their Fourier expansions are written
\begin{eqnarray}
\varepsilon(x,y) &=& \sum_{m,n=-\infty}^{+\infty}\varepsilon_{mn}\,e^{-i(mG_xx+nG_yy)},\\
f(x,y) &=& \sum_{m,n=-\infty}^{+\infty}f_{mn}\,e^{-i(p_mx+q_ny)}.
\end{eqnarray}
We will here derive the matrix expressions of the fundamental relations
\begin{eqnarray}
h(x,y)&=&\varepsilon(x,y)f(x,y),\label{eq:B:fund_h}\\
g_x(x,y)&=&\partial_xf(x,y),\label{eq:B:fund_gx}\\
g_y(x,y)&=&\partial_yf(x,y),\label{eq:B:fund_gy}
\end{eqnarray}
i.e., the relations of multiplication by a function and applying partial derivatives. Assuming the expansions of the new functions
\begin{eqnarray}
h(x,y) &=& \sum_{m,n=-\infty}^{+\infty}h_{mn}\,e^{-i(p_mx+q_ny)},\\
g_x(x,y) &=& \sum_{m,n=-\infty}^{+\infty}g_{x,mn}\,e^{-i(p_mx+q_ny)},\\
g_y(x,y) &=& \sum_{m,n=-\infty}^{+\infty}g_{y,mn}\,e^{-i(p_mx+q_ny)},
\end{eqnarray}
we rewrite Eqs.~(\ref{eq:B:fund_h})--(\ref{eq:B:fund_gy}) using the convolution rule, and applying the partial derivatives as follows:
\begin{eqnarray}
h_{mn} &=& \sum_{k,l=-\infty}^{+\infty}\varepsilon_{m-k,n-l}\,f_{kl},\label{eq:B:fund_h2}\\
g_{x,mn} &=& -ip_m f_{mn},\label{eq:B:fund_gx2}\\
g_{y,mn} &=& -iq_n f_{mn}.\label{eq:B:fund_gy2}
\end{eqnarray}
Assuming furthermore a finite number of the retained Fourier coefficients, i.e., using the summation $\sum_{m=-M}^{+M}\sum_{n=-N}^{+N}$, we can renumber all the indices so that instead of a couple of two sets $m\in\{-M,\ -M+1,\ \ldots,\ M\}$ and $n\in\{-N,\ -N+1,\ \ldots,\ N\}$ we have a single set of indices $\alpha\in\{1,\ 2,\ \ldots,\ \alpha_{\max}\}$, with $\alpha_{\max}=(2M+1)(2N+1)$, related
\begin{eqnarray}
\alpha(m,n) &=& m+M+1+(n+N)(2M+1),\label{eq:B:index_alpha}\\
n(\alpha) &=& (\alpha-1)\mathbf{div}(2M+1)-N,\label{eq:B:index_n}\\
m(\alpha) &=& (\alpha-1)\mathbf{mod}(2M+1)-M,\label{eq:B:index_m}
\end{eqnarray}
where ``$\mathbf{div}$'' denotes the operation of integer division and ``$\mathbf{mod}$'' the remainder (the modulo operation). Then we can rewrite Eqs.~(\ref{eq:B:fund_h2})--(\ref{eq:B:fund_gy2}) into the matrix relations
\begin{eqnarray}
[h] &=& [\![\varepsilon]\!][f],\\
\left[g_x\right] &=& -i\mathbf{p}\,[f],\\
\left[g_y\right] &=& -i\mathbf{q}\,[f],
\end{eqnarray}
where $[f]$, $[h]$, $[g_x]$, and $[g_y]$ are column vectors whose $\alpha$th elements are the Fourier $[m,n]$ elements of the functions $f$, $h$, $g_x$, and $g_y$, indexed by $\alpha(m,n)$ defined in Eq.~(\ref{eq:B:index_alpha}), and where $[\![\varepsilon]\!]$, $\mathbf{p}$, and $\mathbf{q}$ are matrices whose elements are defined
\begin{eqnarray}
[\![\varepsilon]\!]_{\alpha\beta} &=& \varepsilon_{m(\alpha)-m(\beta),n(\alpha)-n(\beta)},\\
p_{\alpha\beta} &=& p_{m(\alpha)}\delta_{\alpha\beta},\\
q_{\alpha\beta} &=& q_{n(\alpha)}\delta_{\alpha\beta},
\end{eqnarray}
where the indices on the right hand parts are defined by Eqs.~(\ref{eq:B:index_n}) and~(\ref{eq:B:index_m}) and where $\delta_{\alpha\beta}$ denotes the Kronecker delta. As a summary we can say that the multiplication by a function is in the reciprocal space represented by the matrix~$[\![\varepsilon]\!]$ (in the sense of the limit $\alpha_{\max}\rightarrow\infty$), which is a generalization of the Toeplitz matrix used for 1D periodicity, and that the partial derivatives are represented by the diagonal matrices~$-i\mathbf{p}$ and~$-i\mathbf{q}$.

\end{document}